# On the Markovian Assumption in Near-Wall Turbulence: The Case of Particle Resuspension

David Ben-Shlomo[1,2], Ronen Berkovich[1,3,*], Eyal Fattal[2,†]

[1]Department of Chemical Engineering, Ben-Gurion University of the Negev, Beer-Sheva 8410501, Israel
[2]Department of Applied Mathematics, Israel Institute for Biological Research, Ness-Ziona 7410001, Israel
[3]Ilze Katz Institute for Nanoscience and Technology, Ben-Gurion University of the Negev, Beer-Sheva 8410501, Israel

Stochastic models of near-wall turbulence commonly rely on the Markovian assumption, despite evidence that coherent structures induce long-lived temporal correlations. Here, we test the validity of this assumption using micron-sized particle resuspension from the viscous sublayer. Analysis of direct numerical simulation (DNS) data reveals that while high- and low-drag events occur with Poissonian statistics, their internal dynamics are strongly persistent, with a Hurst exponent $H \approx 0.84$, indicating intrinsic non-Markovian behavior. We therefore develop a non-Markovian resuspension model based on a fractional Ornstein–Uhlenbeck process, with physical parameters extracted directly from the DNS flow. Comparative simulations show that the empirical success of classical Markovian models arises not from an accurate description of near-wall dynamics, but from their free parameter $C_0$ acting as a phenomenological surrogate for unresolved flow memory. We further identify a critical regime transition controlled by the event decay rate $\lambda$: strong intermittency ($\lambda < 0.2$) invalidates the Markovian approximation, whereas weak intermittency ($\lambda > 0.2$) renders it physically justifiable. These results define quantitative limits on stochastic modeling in near-wall turbulence.

*Introduction* – Turbulence modeling remains a central challenge in modern physics, particularly for wall-bounded flows where strong inhomogeneity and anisotropy complicate universal scaling [1]. While the Kolmogorov-Obukhov 1941 (K41) [2,3] theory provides robust descriptions for homogeneous, isotropic turbulence in the infinite Reynolds number limit, its applicability to the near-wall region, dominated by coherent structures, is limited. Developing a more general theoretical framework to these turbulent regimes is essential for modeling processes like particle resuspension, where the detachment mechanism for particles fully immersed within the viscous sublayer (with dimensionless diameters smaller than five wall units, $d_p^+ < 5$) remains unresolved [4–6].

Two main approaches exist for modeling the driving forces of particle detachment. One attributes detachment to particle-surface interactions, such as collisions with asperities [7–9], while the other attributes detachment to particle-fluid interactions generated by near-wall coherent structures [10–12]. Although Sutherland [13] and Cleaver & Yates [14] initially proposed that turbulent bursts could directly lift particles, later observations showed that detachment is typically preceded by rolling [15–17] complicated this view. Subsequent investigations have produced contrasting results: some experiments suggest that bursting has negligible impact since the associated uplift angles are too small [18], whereas others, including direct numerical simulations (DNS) studies, explicitly link detachment to sweep and ejection events [10–12,19–21]. Nevertheless, theoretical models have struggled to resolve the physical mechanism linking these coherent structures to particle resuspension. As van Hout noted, *"Particularly within the field of particle resuspension, most of the theoretical works seems to neglect the incorporation of the actual physical mechanism linking coherent structures with resuspension"* [19].

Addressing this issue requires accounting for the flow structure. The viscous sublayer ($y^+ < 5$, where $y^+ = \frac{y u_\tau}{\nu}$ denotes the dimensionless wall-normal coordinate, $u_\tau$ is the friction velocity and $\nu$ is the kinematic viscosity) is not laminar but consists of alternating high- and low-velocity streaks exhibiting significant temporal persistence [22–24]. Quadrant analysis [25,26], identifies these coherent motions as ejections (Q2) and sweeps (Q4), which dominate turbulence production and manifest as low- and high-drag events, respectively [27]. Furthermore, recent high-resolution DNS and experimental studies [28–31] confirm that the viscous sublayer hosts rare, intense intermittent events (e.g., backflow) originating from multi-scale interactions. Importantly, the existence of identifiable precursor signatures for these events implies that near-wall fluctuations are not random, independent increments, but organized sequences with distinct temporal correlations.

We hypothesize that this structural intermittency governs the high-order fluctuations responsible for particle detachment. However, existing stochastic models [9,32–34] predominantly rely on the classical K41 theory, treating velocity fluctuations as a Markovian process. This implicitly assumes that velocity increments become uncorrelated over time (i.e., the flow has no "memory"). While this assumption is justified for some free shear flows [35–37], it is questionable in the viscous sublayer, where persistent coherent structures induce strong spatial and temporal correlations. Nevertheless, previous Markovian models often achieve good reconstruction of experimental data by fitting free parameters (associated with particle-surface or particle-flow interactions) to the measurements, potentially masking the underlying mechanism of the process.

*Contact author: berkovir@bgu.ac.il
†Contact author: eyalf@iibr.gov.il

In the present letter, we investigate whether the Markovian assumption is physically justified in the near-wall region. We analyze DNS data to characterize the memory of the flow and introduce a generalized non-Markovian stochastic framework. Consistent with the refined Kolmogorov theory (1962) for intermittent turbulence (K62) [38], this model explicitly links the statistics of particle detachment to the persistent dynamics of coherent structures.

*Theory and methodology* – As introduced above, near-wall coherent structures manifest as high- and low-drag events in the turbulent boundary layer [39,40]. Their persistent nature implies that the forces acting on a particle are not random but strongly time correlated. Accordingly, we focus on characterizing the timescales, amplitudes, and memory properties of these events to link particle motion within the viscous sublayer to the underlying flow topology.

We analyze DNS data from the John Hopkins Turbulent Database (JHTDB) for a fully developed turbulent channel flow at $Re_\tau = \frac{u_\tau h}{\nu} \approx 1{,}000$ (where $h$ is the channel half-height) [41,42]. The simulation domain covers $8\pi h \times 2h \times 3\pi h$ in the streamwise, $x$, wall-normal, $y$, and spanwise, $z$ directions, respectively, and fully resolves scales from energy-containing eddies down to the dissipative Kolmogorov scales. Given that the stochastic resuspension framework is formulated as a single-particle model, the particle concentration is assumed to be sufficiently dilute so that the particles do not interact with each other or significantly alter the flow. Under these conditions, the particles respond to the turbulent velocity field without feeding back onto it, consistent with the one-way coupling assumption [43,44]. This is supported by findings that particle and fluid velocity fluctuations in the sublayer are of comparable magnitude, implying minimal feedback [45]. Thus, the DNS velocity field is used as a prescribed input.

We identify events using an amplitude criterion based on the normalized wall shear stress (WSS), defined as the ratio of the instantaneous WSS value to its time-averaged value, $\tau_w/\langle\tau_w\rangle$. An amplitude criterion is adopted, where events are classified as high- or low-drag when the normalized WSS exceeded a ± 5% threshold. This value was chosen over larger thresholds (e.g., ± 10%) to increase the number of detected events, thereby improving the statistical robustness of the analysis [46,47]. A recent research demonstrated that adopting inner scaling ($\Delta t^+ = \Delta t u_\tau^2/\nu$) renders the fraction of time spent in high- and low-drag events independent of the friction Reynolds number $Re_\tau$ [27]. Based on this finding, we employed inner time scaling in our analysis. No minimum-duration criterion is imposed, consistent with prior study [47], due to the short characteristic timescales of events in the viscous sublayer. To quantify memory effects in the WSS signal, we estimate the Hurst exponent ($H$) via the Rescaled Range (R/S) analysis. This method was adopted over other alternatives (e.g., Detrended Fluctuation Analysis) [48–50], due to its extensive use and demonstrated robustness, [50,51].

Our non-Markovian generalization builds upon our previously introduced Lagrangian stochastic Markovian model, which demonstrated good agreement with experimental data [32]. The model assumes a particle that rolls in a single direction [4,15,52,53], with its motion governed by a moment balance, $M$, about a pivot point:

$$M \equiv I\frac{d\omega}{dt} = bF_D + \frac{a}{2}F_L - r_a F_{adh} - \frac{a}{2}mg \qquad (1)$$

where $I$ is the particle moment of inertia, approximated as a sphere and evaluated using the parallel axis theorem as $I = \frac{7}{20}md_p^2$. The particle diameter and mass are denoted by $d_p$ and $m$, respectively. $g$ is the gravitational acceleration, $\omega$ is the particle angular velocity, $a$ is the distance between consecutive surface asperities, $r_a$ is the lever arm of the adhesion force, and $b = \sqrt{\left(\frac{d_p}{2}\right)^2 - \left(\frac{a}{2}\right)^2}$. The hydrodynamic forces $F_D$ (drag) and $F_L$ (lift) are modeled using Stokes drag [54] with O'Neill's wall-correction [55], and the formulation of Mollinger & Nieuwstadt [56] combined with Maude's wall-correction factor [57], respectively. The mean adhesion force $F_{adh}$ is calculated using the Rabinovich approximation [58], which describes the interaction potential between a spherical particle and a rough surface.

Within this framework, the overall angular velocity $\omega$ is separated into mean and fluctuating components. The mean component is determined by the forces in Eq. (1), while the fluctuating component, $\omega'$, is modeled as a stochastic process. Consistent with K41 theory [45], this model uses a standard Ornstein-Uhlenbeck process (OUP) [59,60]:

$$d\omega'(t) = -\omega'(t)\frac{dt}{T} + \sqrt{2\sigma^2/T}\,dW(t) \qquad (2)$$

where $T$ is the integral timescale, $\sigma$ is a parameter that controls the variance and the autocorrelation of the process, and $W(t)$ denotes a Wiener process [59]. When the particle's kinetic energy exceeds a critical threshold, such that $\omega > \omega_c = 2a/(d_p\Delta t)$ ($\Delta t$ is the simulation time-step), the surface adhesion energy is overcome and the particle resuspends.

It is important to highlight that according to Fu et al. [33], the parameter $\sigma$ depends on a free model constant, $C_0$, whose precise physical interpretation is unclear but is generally associated with the integral time scale of the fluctuating moment. Based on our hypothesis, we assume that the physical meaning of this constant reflects the intermittent dynamics in the viscous sublayer, induced by near-wall coherent structures.

The observed penetration of intermittent coherent structures into the viscous sublayer seem to contradict the assumption of uncorrelated white noise, as it may not adequately capture the inhomogeneous dynamics of near-wall turbulence. Faranda et al. [61] demonstrated that Hurst exponents provide a useful measure of deviations from the K41-based Markovian description in inhomogeneous and anisotropic turbulent flows, linking these deviations to the presence of intermittent coherent structures. Motivated by this insight, we propose a generalized non-Markovian model in which the standard OUP is replaced with a fractional Ornstein–Uhlenbeck process (fOUP) [62]. This formulation is consistent with the refined K62 theory [38] and is obtained by replacing the Wiener process increment $dW(t)$ in Eq. (2) with an increment of fractional Brownian motion (fBm), $dB^H(t)$. The governing stochastic differential equation for the fluctuating component becomes

$$d\omega'(t) = -\omega'(t)\frac{dt}{T} + \sqrt{2\sigma^2/T}\,dB^H(t) \qquad (3)$$

Here, the Hurst exponent $H$ (estimated from DNS data in the results section) governs the memory of the process. When $H$ = 0.5, the formulation reduces to the standard Wiener process. For $0.5 < H < 1$, the process becomes persistent, meaning that fluctuations tends to continue in the same direction over time, thus capturing the influence of coherent structures [61,63–65]. The fBm can be represented as a cumulative integral of a Gaussian random process:

$$B^H(t) = \frac{1}{\Gamma\left(H+\frac{1}{2}\right)}\int_{-\infty}^{0}\left[(t-s)^{H-\frac{1}{2}}-(-s)^{H-\frac{1}{2}}\right]dW(s) + \frac{1}{\Gamma\left(H+\frac{1}{2}\right)}\int_{0}^{t}(t-s)^{H-\frac{1}{2}}dW(s) \qquad (4)$$

where $\Gamma(H)$ is the gamma function [66]. We further note that, although the fOUP is a generalized form of the OUP by incorporating non-Markovian memory effects, the one-point statistics of both processes satisfy the same general form of the Fokker-Planck equation [67]:

$$\frac{\partial}{\partial t}P(\omega',t|\omega'_0,t_0) = \left(-\frac{\partial}{\partial\omega'}D_1(\omega',t) + \frac{\partial^2}{\partial\omega'^2}D_2(\omega',t)\right)P(\omega',t|\omega'_0,t_0) \qquad (5)$$

where $P(\omega',t|\omega'_0,t_0)$ is the conditional probability density function (PDF) (with $t > t_0$), $D_1(\omega',t)$ is the drift coefficient and $D_2(\omega',t)$ is the dispersion coefficient. Although the structural form of Eq. (5) is identical for both models, memory effects in the fOUP enter through a time-dependent dispersion coefficient, $D_2 \sim t^{2H-1}$, contrasting with the constant dispersion of the standard OUP. For additional details on the effect of the Hurst parameter on particle dynamics and the numerical algorithm of the non-Markovian model see the supplemental material [68].

*Results and discussion* – A DNS-based analysis was conducted to characterize the spatio-temporal properties of high- and low-drag events within the viscous sublayer. Figure 1 presents the normalized fluctuating WSS as a function of inner-scaled time. Data were sampled at the center of the channel in the streamwise and spanwise directions, specifically (4π, 1.5π), at $0.5 < y^+ < 1.5$ in the wall-normal direction, corresponding to the lower portion of the viscous sublayer [69]. At each time where the normalized WSS exceeded the ± 5% threshold relative to the mean, a high- or low-drag event was identified and recorded for subsequent analysis. Figure 1 shows that the viscous sublayer exhibits substantial shear fluctuations, reaching deviations up to 10% from the mean, in agreement with previous observations [70,71]. These excursions are not merely random noise, as they represent the imprint of persistent coherent structures that penetrate into the viscous sublayer [31,72,73]. The identification of discrete high- and low-drag events (marked in red and blue) confirms that the particle experiences the flow as a sequence of distinct forcing intervals. Collectively, these results support the adoption of a time-correlated, non-Markovian modeling framework to capture the near-wall flow dynamics in the viscous sublayer.

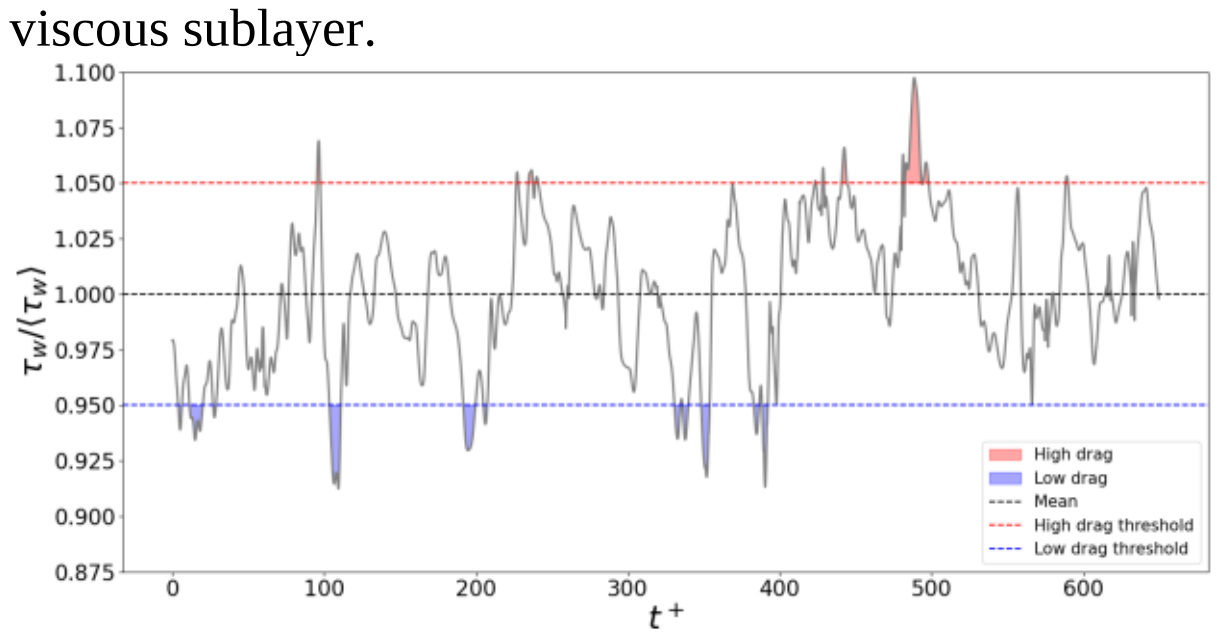

FIG 1. Normalized fluctuating WSS at the center of the channel in the streamwise and spanwise directions (4π, 1.5π), as a function of inner-scaled time. Data were sampled within the viscous sublayer $0.5 < y^+ < 1.5$. High-drag events (red) and low-drag events (blue) are whenever the threshold criterion is exceeded.

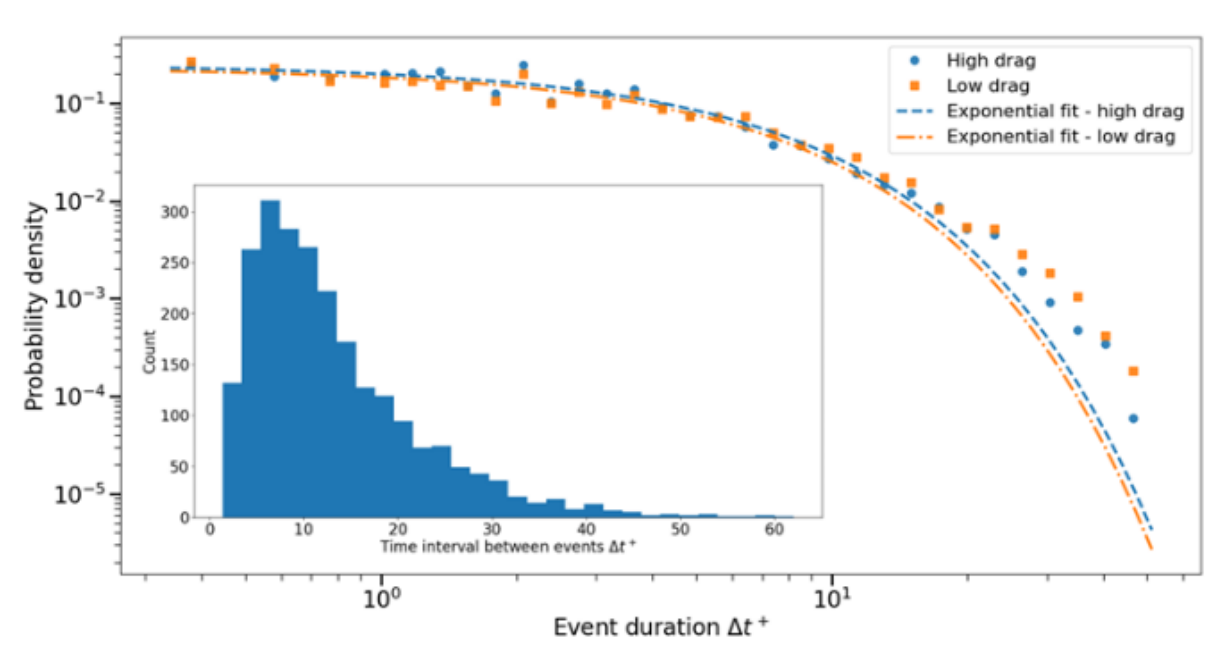

FIG 2. PDFs of high-drag (blue circles) and low-drag (orange squares) events durations within the viscous sublayer, with corresponding exponential fits shown for high-drag events (blue dashed line) and low-drag events (orange dash-dotted line). The inset shows a histogram of the time interval between consecutive events. Both PDFs and histograms were obtained from analyses at 100 distinct spatial locations in the streamwise-spanwise plane within $0.5 < y^+ < 1.5$.

We next calculate the PDF of the duration and occurrence frequency of high- and low-drag events by extending the analysis performed in Fig. 1 to 100 distinct streamwise-spanwise locations within the viscous sublayer. Figure 2 presents the resulting PDFs of event durations, and its inset shows the histogram of the time intervals between consecutive events. Each distribution can be fitted with an exponential function of the form $P(\Delta t^+) = Ae^{-\lambda\Delta t^+}$, where $A$ is the amplitude and $\lambda$ is the decay rate. We note that since the time is expressed in inner-scaled, dimensionless units, the resulting decay rate is likewise dimensionless. As illustrated in Fig. 2, both PDFs are well described by a single exponential. This behavior is characteristic of a Poisson process, implying that the termination of a drag event is a random, memoryless occurrence. However, a crucial distinction should be made: while the switching between events follows Poissonian statistics (random occurrence), the dynamics during an event are governed by the coherent structure itself, and therefore exhibit strong temporal correlation. Table 1 summarizes the exponential fit parameters.

Notably, the decay rates obtained here ($\lambda \approx 0.2$) are approximately an order of magnitude larger than those reported by Agrawal et al. [27]. This difference likely stems from the differing event-selection threshold criteria. Agrawal et al. [27] imposed a strict minimum time-duration filter ($t^+ > 200$), effectively isolating the larger structures, whereas our approach captures the full spectrum of viscous-sublayer events relevant to the short timescales of particle resuspension. The histogram of the time intervals between consecutive events (Fig. 2 inset), where high- and low-drag events are counted together is based on 2,354 detected events (counts). The median interval is $10.8\Delta t^+$ and the mean interval is $13.2\Delta t^+$. These findings are consistent with previous numerical and experimental studies, which reported that the average spacing between near-wall velocity spikes is typically less than $20\Delta t^+$ [74,75].

TABLE I. Amplitude ($A$) and decay rate ($\lambda$) of high- and low-drag events within the viscous sublayer at $Re_\tau \approx$1,000, including their standard error. The goodness of fit ($R^2$) to the exponential function is indicated in brackets next to each event type.

| Event type | $A$ | $\lambda$ |
|---|---|---|
| High drag (0.93) | 0.223±0.011 | 0.203±0.022 |
| Low-drag (0.90) | 0.239±0.014 | 0.239±0.030 |

To investigate the non-Markovian nature of the viscous sublayer dynamics, we perform a time-correlation analysis to quantify the degree of temporal persistence and extract the Hurst exponent from the DNS data. This was accomplished using the R/S analysis (described in the theory section) that applied to the temporal evolution of the wall-normal shear stress, $\tau_w(t)$, sampled within $y^+ \in (0.5, 1.5)$. The sample length, $n$, defines a temporal window over which the near-wall turbulent fluctuations are evaluated. For each block of length $n$, the adjusted range $R$ is computed as the difference between the maximum and minimum of the cumulative deviation from the block mean, capturing the largest persistent excursion of the shear stress relative to the local average. The standard deviation $S$ represents the root-mean-square magnitude of the fluctuations within the same window, reflecting the local turbulent intensity. The ratio $R/S$ thus provides a normalized measure of persistence relative to fluctuation strength, and its scaling with $n$ yields the Hurst exponent $H$, which characterizes the temporal correlation and memory in near-wall turbulence.

Figure 3 portrays the Hurst exponent estimation from DNS WSS time series using R/S analysis. Figure 3(a) presents a representative R/S calculation for a single spatial location (channel center), while Fig. 3(b) displays a heat map of Hurst values computed at 100 spatial locations across the channel. The Hurst exponent varies between 0.79 and 0.87 with a median $H_{med}$ = 0.84 and a mean $\overline{H}$ = 0.83. These values indicate pronounced long-range temporal correlations and strong persistence in the near-WSS signal, supporting a non-Markovian description of the turbulent fluctuations in the viscous sublayer.

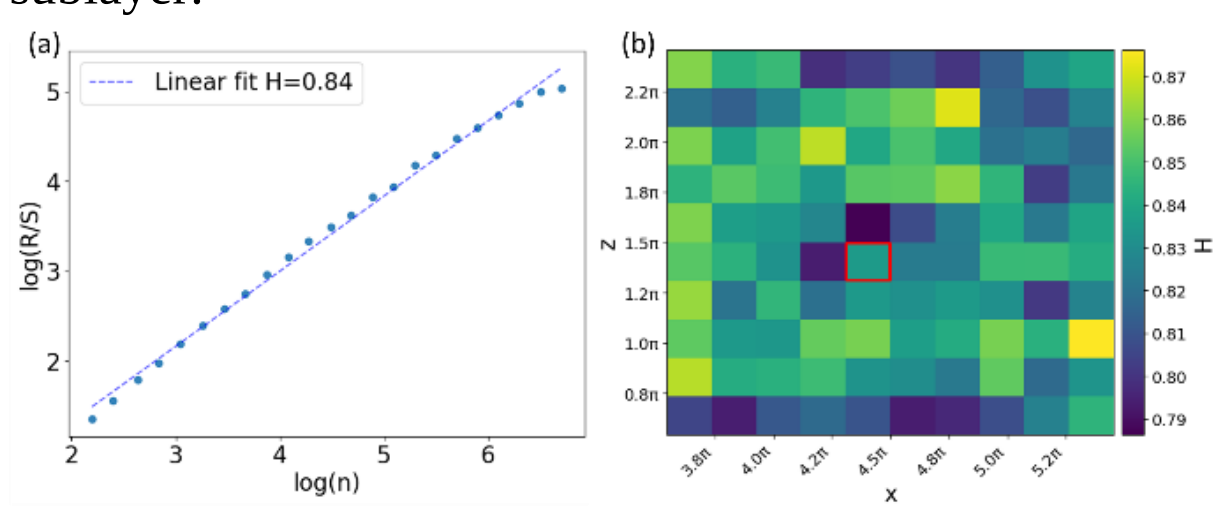

FIG 3. Estimation of the Hurst exponent using rescaled range analysis. (a) Example for a single point located at the channel center in the streamwise and spanwise directions (4π, 1.5π) and within $0.5 < y^+ < 1.5$. According to the R/S method, the slope of the linear fit yields a Hurst exponent of $H$ = 0.84. (b) Heat map of Hurst exponent values computed from 100 spatial locations across the channel. The data point shown in (a) is indicated by a red square outline in (b).

Following the characterization of flow dynamics within the viscous sublayer and the integration of extracted timescales and correlation intensities into the non-Markovian resuspension model, we address a fundamental question: given that the viscous sublayer is inherently non-Markovian ($H \approx 0.84$), why do classical Markovian resuspension models [9,32–34] often reproduce experimental data successfully? To resolve this, we investigate whether the free parameter in the Markovian model, $C_0$, effectively compensates for the missing physics associated with flow memory.

We perform comparative numerical simulations using both the non-Markovian and Markovian frameworks, with parameters that characterize the tracking of tungsten particles ($d_p$ = 13 μm) on a hydrodynamically smooth surface (with root mean square roughness of 2.2 nm), ensuring full immersion in the viscous sublayer. All additional model parameters are specified in [32]. In the non-Markovian model, the event decay rate $\lambda$ was varied to probe the sensitivity of the system to intermittency, while in the Markovian model, the parameter $C_0$ which scales the stochastic forcing variance and is theoretically linked to the integral timescale of turbulent fluctuations, was systematically varied.

Figure 4(a) illustrates the sensitivity of the Markovian model to $C_0$. As $C_0 \to 0$, the stochastic contribution vanishes, and resuspension is driven solely by mean forces, resulting in systematic underprediction. Crucially, we observe that when $C_0 = 1 \times 10^{-3}$, the specific value empirically calibrated by Fu et al. [33] to match experimental measurements, the Markovian prediction aligns with our non-Markovian model (with $\lambda = 0.2$). This agreement reveals an important physical insight: under viscous sublayer conditions, the non-Markovian and Markovian formulations can produce equivalent macroscopic predictions, but through fundamentally different mechanisms. The non-Markovian model achieves this by explicitly incorporating the structure and memory of drag events through the $H$, and $\lambda$ parameters, whereas the Markovian model achieves similar outcomes by tuning the empirical parameter $C_0$. We therefore propose that $C_0$ acts not merely as a fitting parameter, but as a phenomenological surrogate for the integrated memory of high- and low-drag events. In effect, by tuning $C_0$, previous studies effectively adjusted the intensity of the white-noise forcing to emulate the persistence imparted by coherent structures.

To delineate the limits of this equivalence, Fig. 4(b) examines the effect of the event decay rate, $\lambda$, on resuspension. We identify a distinct regime transition around a critical threshold of $\lambda \approx 0.2$. In the strong intermittency regime ($\lambda < 0.2$), where events are long-lived, the resuspension profile exhibits a pronounced plateau at low friction velocities ($u_\tau < 1$ m/s). This regime is strongly non-Markovian, as the flow memory persists long enough for the particle to experience sustained forcing, leading to dynamics that cannot be reproduced by the white-noise approximation used in Fig. 4(a). Importantly, the Markovian model fails to capture the characteristic plateau observed in this regime, regardless of the calibration of $C_0$. Conversely, in the weak intermittency regime ($\lambda > 0.2$), events become shorter and more frequent, causing the plateau to disappear and the curve to smooth out. In this limit, coherent structures decay so rapidly that they appear as effectively uncorrelated fluctuations from the particle's perspective, rendering the Markovian approximation physically justified. This aligns with observations in inhomogeneous canopy flows, where rapid velocity decorrelation timescale justifies a Markovian description [76,77].

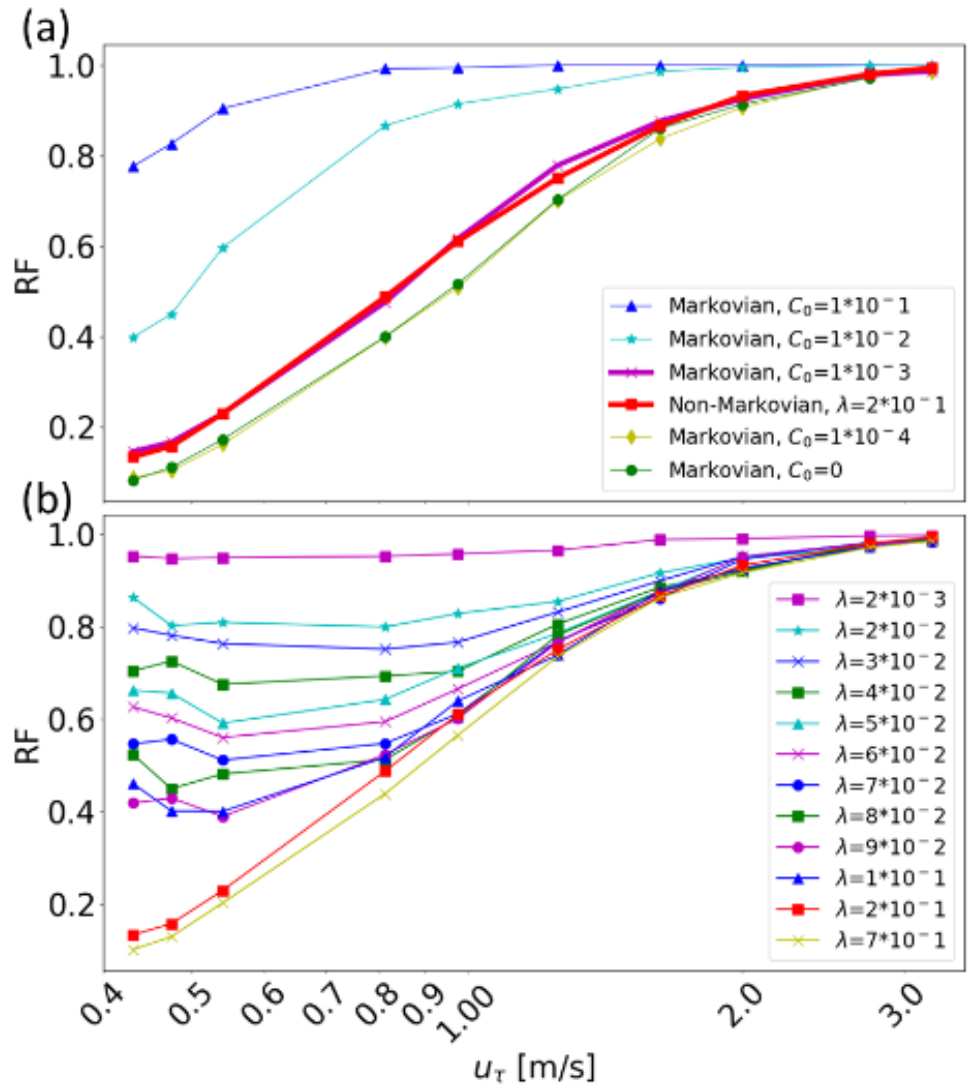

FIG 4. Resuspension fraction (RF) as a function of friction velocity for particles fully immersed within the viscous sublayer. (a) Comparison of the Markovian model (varying $C_0$) with the non-Markovian model (at $\lambda =$ 0.2). (b) Non-Markovian model predictions for varying high- and low-drag events decay rates, $\lambda$.

We emphasize that the validity of the Markovian approximation depends on the specific imposed boundary conditions. In the present analysis we assumed a hydrodynamically smooth wall, however, surface roughness is expected to reduce the sublayer thickness [78], potentially altering the frequency and dynamics of coherent structures. Furthermore, the assumption of a one-way coupling regime neglects particle-flow feedback mechanisms where, at higher concentrations, particles intensify sweep and ejection events [79], creating a coupled system likely requiring non-Markovian descriptions. Finally, for processes occurring higher in the boundary layer, such as particle deposition and sedimentation, the influence of large-scale structures becomes more pronounced, consistent with our calculation of a decay rate ($\lambda \approx 0.1$) at the buffer layer ($y^+$ = 50). Recent work by Hung and Tsai [63] demonstrated that a non-Markovian model outperforms a Markovian formulation for reconstructing sedimentation measurements, utilizing a decay rate ($\lambda \approx$ 0.02) consistent with the stronger intermittency of the buffer layer. Thus, while the Markovian approximation can be tuned to describe simple sublayer resuspension, a non-Markovian framework offers a more general and physically grounded description that is essential for flows dominated by memory effects.

*Acknowledgments* – The authors wish to thank Ziv Klausner and Yehuda Arav for helpful discussions. The authors of this work are grateful for the generous support of Pazy Foundation, grant ID 133-2020 and the John R. Goldsmith Memorial Prize Fund.

*Data availability* – The data that support the findings of this article are not publicly available. The data are available from the authors upon reasonable request.

# Supplemental material

for

# On the Markovian assumption in near-wall turbulence: The case of particle resuspension

David Ben-Shlomo[1,2], Ronen Berkovich[1,3,*], Eyal Fattal[2,*]

[1]Department of Chemical Engineering, Ben-Gurion University of the Negev, Beer-Sheva 8410501, Israel
[2]Department of Applied Mathematics, Israel Institute for Biological Research, Ness-Ziona 7410001, Israel
[3]Ilze Katz Institute for Nanoscience and Technology, Ben-Gurion University of the Negev, Beer-Sheva 8410501, Israel

## 1. Influence of the Hurst exponent on particle dynamics

The fractional Brownian motion (fBm) generalizes classical Brownian motion by allowing correlated increments. It is characterized by a zero mean $\langle B^H(t)\rangle = 0$, a variance $\langle \left(B^H(t)\right)^2\rangle = t^{2H}$, and a covariance $\langle B^H(s)B^H(t)\rangle = \frac{1}{2}\left(|s|^{2H} + |t|^{2H} - |t-s|^{2H}\right)$. The fBm can be represented as a cumulative integral of a Gaussian random process:

$$B^H(t) = \frac{1}{\Gamma\left(H+\frac{1}{2}\right)}\int_{-\infty}^{0}\left[(t-s)^{H-\frac{1}{2}} - (-s)^{H-\frac{1}{2}}\right]dW(s) + \frac{1}{\Gamma\left(H+\frac{1}{2}\right)}\int_{0}^{t}(t-s)^{H-\frac{1}{2}}dW(s) \qquad (S1)$$

where $\Gamma(H)$ is the gamma function [1]. The exponent $H$ is a real number in (0, 1) and governs the memory properties of the process. Specifically, for $H$ = 0.5, the standard Wiener process is recovered. For 0 < $H$ < 0.5 the process exhibits anti-persistence, meaning that increases are likely to be followed by decreases and vice versa. Conversely, for 0.5 < $H$ < 1, the process is persistent, with trends tending to continue in the same direction. The rescaled range (R/S) method, developed by Hurst [2], is expressed as

$$log(R/S) = log(c) + Hlog(n) \qquad (S2)$$

where *R* is the adjusted range, *S* is the standard deviation, *c* is a constant, *H* is the Hurst exponent and *n* is the sample length. Equation (S2) is linear in its current form, allowing the Hurst exponent *H* to be obtained through a linear regression of the data.

To illustrate the influence of the Hurst exponent on particle dynamics, figure S1 presents a comparison of 1,000 sample paths and their correlation generated for various values of *H*, obtained by numerically solving equation (S1) using the Cholesky decomposition method [3]. The trajectories are plotted together with their ensemble mean and standard deviation. As expected from their definitions, both the Wiener process ($H$ = 0.5) and fBm processes exhibit zero ensemble mean. Additionally, as the Hurst exponent increases, the trajectories display greater spatial extent. For $H > 0.5$, the sample paths exhibit positively correlated, smoother behavior with enhanced persistence, enabling them to traverse significantly larger distances due to their dependent increments. Notably, the difference in magnitude between the Wiener ($H$ = 0.5) and fBm with $H$ = 0.8 processes can reach up to an order of magnitude. These characteristics are also emphasized in the normalized covariance by the product of standard deviations of each *H* value. In case of the Wiener process ($H$ = 0.5), the normalized covariance structure reduces to the Brownian case, which decays proportionally to the smaller time. For persistent behavior ($H > 0.5$), The normalized covariance remains high even far off the diagonal, indicating long memory. For anti-persistent behavior ($H < 0.5$), the normalized covariance decays quickly off the diagonal, values at distant times become nearly uncorrelated.

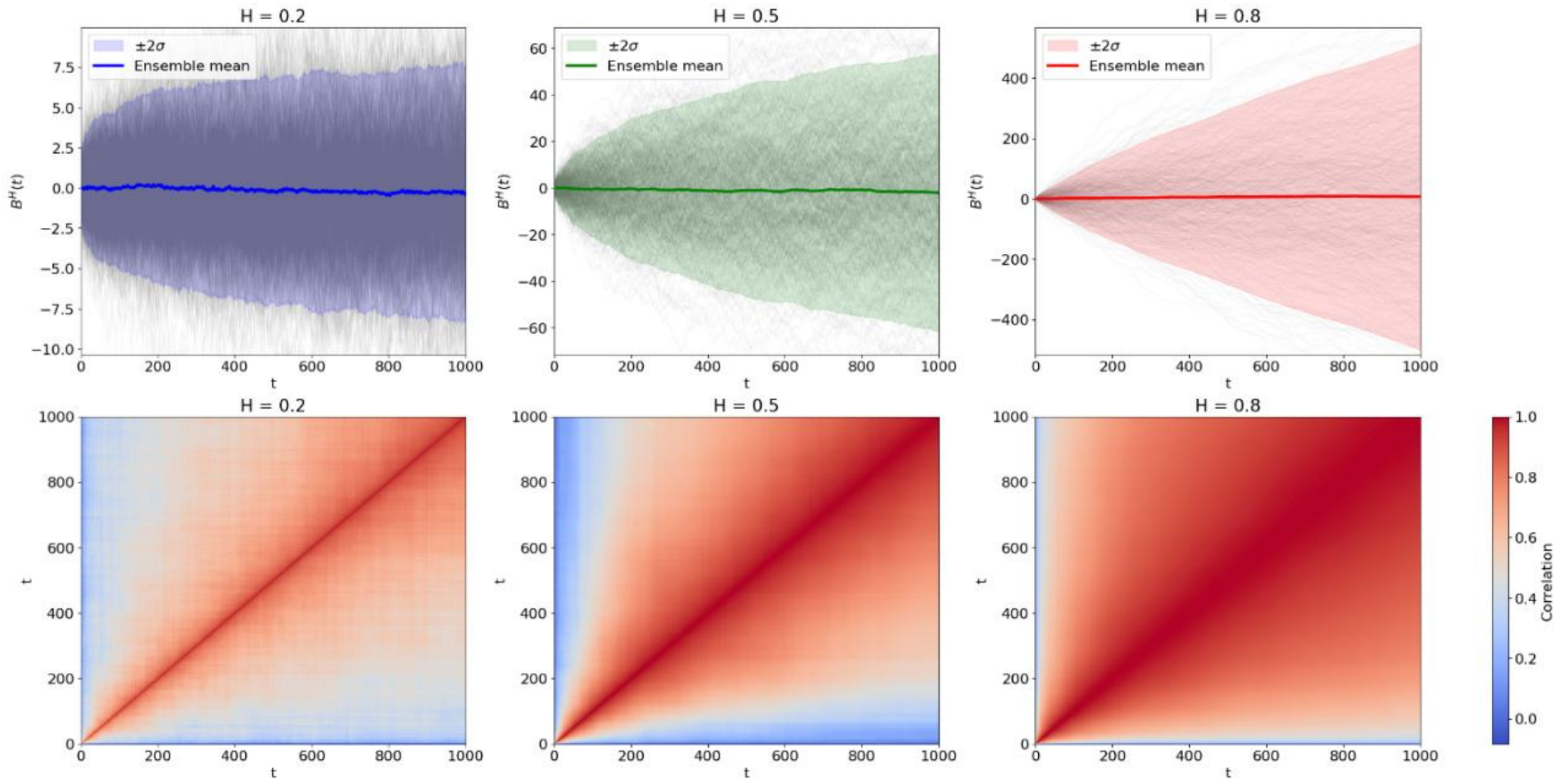

**Fig. S1.** Top: sample paths (trajectories) of fBm for three different Hurst exponents: *H* = 0.2 (blue) *H* = 0.5 (green) *H* = 0.8 (red), based on 1,000 realizations each. Bottom: correlation heat map corresponding for the different Hurst exponent values. The correlation here is the normalized covariance by the product of standard deviations. Both the time variable *t* and the fBm values $B^H(t)$ are dimensionless.

## 2. Numerical simulation procedure

The non-Markovian simulation proceeds as follows. At the start of the simulation, the particle is set to rest on the surface. At each subsequent time step, $\Delta t$, the particle may encounter a high- or low-drag event associated with turbulent coherent structures. The probability of encountering such an event is determined from the event frequency analysis performed on direct numerical simulation (DNS) data. If the particle encounters an event, the fluctuating angular velocity is computed using the fractional Ornstein-Uhlenbeck process (fOUP), with both the Hurst exponent and the event duration derived from the DNS-based analysis conducted in this study. In the absence of an event, the fluctuating velocity is calculated using the classical Ornstein-Uhlenbeck process (OUP), consistent with Markovian dynamics. The total angular velocity of the particle, obtained as the sum of the mean and fluctuating contributions following Reynolds decomposition, is then compared to the critical resuspension threshold, $\omega_c$. If the total angular velocity exceeds the critical value, the particle is considered to overcome the adhesive forces and is assumed to be resuspended. Otherwise, the particle remains on the surface, and the simulation proceeds to the next time step, repeating the same procedure iteratively.

The integration of both the OUP and fOUP was performed using the Euler scheme, which has been previously demonstrated to be a simple and reliable method for solving stochastic differential equations [4,5]. It is further emphasized that, since the stochastic term in the Langevin equation is independent of the solved variable ($\omega'$ in this case), the integration of the stochastic differential equation is straightforward and the Itô-Stratonovich dilemma does not apply [6]. The simulation time step was selected to be equal to the turbulent integral timescale, $T_0 = \nu/{u_\tau}^2$, which is two orders of magnitude smaller than the average bursting interval ($300\nu/u_\tau^2$) [7] and one order smaller than the integral timescale within the viscous sublayer ($20\nu/u_\tau^2$) for the streamwise wall shear stress [8], ensuring adequate temporal resolution to sample the memory effect. The total simulation duration was set for one second, reflecting the short timescale of the resuspension process, which occurs on the order of milliseconds [4]. Each simulation tracks 1,000 particles, a number determined through sensitivity analyses showing that increasing the particle count does not alter the predicted resuspended fraction, thereby confirming statistical significance. Other simulation parameters remain consistent with those defined in [4,9]. Finally, owing to the limited variation of the Hurst exponent obtained from the rescaled range analysis, and the demonstrated insensitivity of the resuspension results to this variation, the single representative value $H$ = 0.84 (the median from the analysis) was adopted in the numerical algorithm of the non-Markovian model.

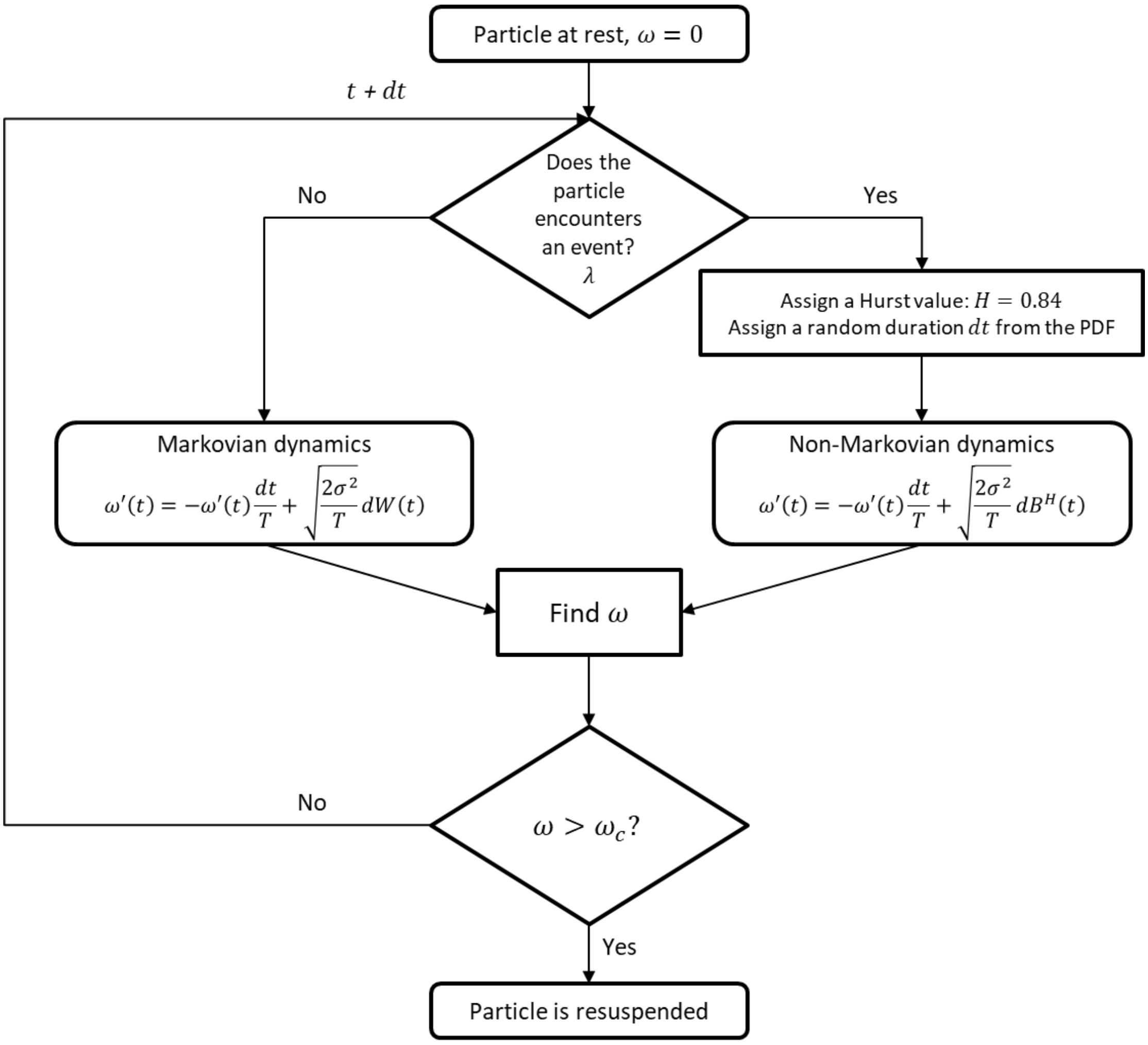

**Fig. S2.** Simulation algorithm flowchart.